\documentclass[12pt,twosided]{article}
\usepackage[latin1]{inputenc}
\usepackage{amsmath, graphicx, natbib, layout, verbatim}
\usepackage{setspace, amssymb}
\usepackage{fancyhdr, mathrsfs}
\usepackage{tikz}
\usetikzlibrary{arrows,matrix}
\DeclareGraphicsExtensions{.eps}
\usepackage[hmargin=2cm,vmargin=3.5cm]{geometry}

\newenvironment{frcseries}{\fontfamily{frc}\selectfont}{}

\title{A Generalized Fellegi--Sunter Framework for Multiple Record Linkage With Application to Homicide Record--Systems}

\author{
Mauricio Sadinle and Stephen E. Fienberg\\\\
\textsc{Carnegie Mellon University, Pittsburgh, PA 15213-3890} }
\date{}

\begin{document}
\maketitle

\begin{center}
\begin{minipage}[b]{0.85\textwidth}
\footnotesize{Mauricio Sadinle is a Ph.D. student, Department of Statistics, Carnegie Mellon University, Pittsburgh, PA 15213 (e-mail: msadinle@stat.cmu.edu); and Stephen E. Fienberg is Maurice Falk University Professor of Statistics and Social Science in the Department of Statistics, the Machine Learning Department, and the Heinz College, Carnegie Mellon University (e-mail: fienberg@stat.cmu.edu).  This research was partially supported by NSF Grants BCS-0941518 and SES-1130706 to Carnegie Mellon University, and by the Singapore National Research Foundation under its  International Research Centre @ Singapore Funding Initiative and administered by the IDM Programme Office.  The authors thank Rob Hall, Kristian Lum, Michael Larsen, the Associate Editor and two referees for helpful comments and suggestions on earlier versions of this paper, and Jorge A. Restrepo for providing the Colombian homicide data.    An early version of this paper was written by the first author when he was affiliated to the Conflict Analysis Resource Center (CERAC) and the National University of Colombia at Bogot\'a.}
\end{minipage}
\end{center}
\abstract{We present a probabilistic method for linking multiple datafiles.  This task is not trivial in the absence of unique
identifiers for the individuals recorded.  This is a common scenario when linking census data to coverage measurement surveys for census coverage evaluation, and in general when multiple record--systems need to be integrated for posterior analysis.  Our method generalizes the Fellegi--Sunter theory for linking records from
two datafiles and its modern implementations.
The goal of multiple record linkage  is to classify the record $K$-tuples coming from $K$ datafiles according to the different matching patterns.
Our method incorporates the transitivity of
agreement in the computation of the data used to model matching
probabilities.  We use a mixture
model to fit matching probabilities via maximum likelihood using the EM algorithm. We present a method to
decide the record $K$-tuples membership to the subsets of matching patterns and we prove its optimality.   We  apply our method to the integration
of the three Colombian homicide record systems and  perform a simulation study to explore the performance of the method under measurement error and different scenarios.  The proposed method works well and opens new directions for future research.
}

\section{INTRODUCTION}

Record linkage is a widely--used technique for identifying records that refer to the same individual across different datafiles. This task is not trivial when unique identifiers are not available, and many
authors have proposed probabilistic methods to deal with
this problem building upon the seminal work of \citet{Newcombeetal59} and
\citet{FellegiSunter69}.  Applications of record linkage include merging post--enumeration surveys and census data for census coverage evaluation \citep[e.g.,][]{Winkler88,Jaro89,WinklerThibaudeau91}, linking health--care databases for epidemiological studies \citep[e.g.,][]{Belletal94,Merayetal07}, and adaptive name matching in information integration \citep{Bilenkoetal03} among others.

\subsection{Linking Multiple Datafiles}

To perform record linkage involving more than two datafiles,
some authors have used record linkages for each pair of datafiles or other ad hoc procedures
\citep[e.g., see][]{Darrochetal93, ZaslavskyWolfgang93, AsherFienberg01, Asheretal03, Merayetal07}.  Separate pairwise matchings of datafiles do not guarantee the
transitivity of the linkage decisions and thus require
resolving discrepancies \citep{FienbergManrique09}.  For
example, let us suppose we link the record of the individual
$a$ in a first datafile and the record of an individual $b$ in a
second datafile from a bipartite record linkage (classical record linkage
of two datafiles). Then, from a second bipartite record linkage, we
link the record of $b$ to the record of an individual $c$ in a third
datafile. Based on these two linkages we might conclude that $a$, $b$, and $c$ are the
same individual.  Unfortunately, had we also linked the first and third files,  $a$ and $c$ may not match.  If $a$, $b$, and $c$  truly correspond to
the same individual, the non--match could occur due to measurement error
or incomplete record information. On the other hand, if the
records of $a$, $b$, and $c$ do not refer to the same individual, we have four possibilities: $a$ and $b$ refer to the same
individual but $c$ refers to another one, $a$ and $c$ refer to the
same individual but $b$ refers to another one, $b$ and $c$ refer to
the same individual but $a$ refers to another one, or all $a$, $b$,
and $c$ refer to different individuals. By using bipartite record
linkage for each pair of files we cannot resolve the matching pattern for these
three records.  While there are various ad hoc approaches to resolve the results of multiple bipartite matchings, no formal methodology has appeared in the statistical literature (e.g., see the recent surveys of Herzog, Scheuren and Winkler 2007, 2010).

\subsection{Census and Record--Systems Coverage Evaluation}\label{ss:recordsystems}

Implementation of accurate methods for census coverage evaluation and
possibly census adjustment requires the integration of multiple datafiles.  The usual methodology of census coverage
evaluation matches a coverage measurement survey to the census data in
order to estimate population sizes using dual-system estimation
\citep{Hogan92,Hogan93}.  This procedure  is subject to
``correlation bias,'' which results when responses to the census and
survey are dependent or the joint inclusion probabilities are
heterogeneous
\citep{Darrochetal93,ZaslavskyWolfgang93,AndersonFienbergbook}.  The
incorporation of additional surveys or administrative data into the
coverage evaluation process allows for checking on assumptions regarding
independence of lists and homogeneity, and for modeling departures from
them.   This in turn requires attention to the problem of multiple
record linkage.

Likewise, under--registration is the norm rather than the exception in record--systems of human rights violations and violent events in general, especially in countries with high levels of violence.  Discrepancies appear whenever there are different record--systems capturing information about the same event of interest.  The diversity of sources provides a useful input for coverage assessment of the different record--systems \citep[e.g.,][]{Ball00,Gohdes10,Lumetal10}.  A clear example of this scenario occurs in Colombia, where there exist three homicide record--systems which usually differ in the number of recorded casualties.  Those record--systems are maintained by the Colombian Census Bureau
(Departamento Administrativo Nacional de Estadistica -- DANE, in Spanish), the Colombian National
Police (Policia Nacional de Colombia), and the Colombian Forensics Institute
(Instituto Nacional de Medicina Legal y Ciencias Forenses).  The discrepancies in the numbers recorded by these record--systems are the result of conceptual and methodological differences among these institutions, as well as problems of geographical coverage \citep{RestrepoAguirre07}.  Whereas the data from the National Police and Forensics Institute simply record the information obtained from their daily activities, the objective of the Colombian Census Bureau is to determine the true number of deaths occurring in Colombia and its geographical subdivisions \citep{DANEvitales}.  Thus, the coverage evaluation of the Colombian Census Bureau record--system is important, and its linkage with the other two sources can lead to improved estimates of the number of homicides.

\subsection{Overview of the Article}

We propose a method for the
linkage of multiple datafiles, generalizing the theory of
\citet{FellegiSunter69} and the implementations presented by
\citet{Winkler88} and \citet{Jaro89}, which still represent  the mainstream approach for unsupervised record linkage (see \citet{CopasHilton90} for a supervised approach).  Our method
incorporates the transitivity of agreement in the computation of the data used to
model matching probabilities. In Section \ref{s:subsets}
we generalize the set of record pairs presented by
\citet{FellegiSunter69} to a $K$-ary product of the $K$ datafiles to
be linked, and we present this $K$-ary product as the union of all
the possible subsets that contain the possible patterns of agreement
of the record $K$-tuples. In Section \ref{s:data} we propose a
method to compute comparison data from
record $K$-tuples, incorporating transitivity, and we present a way to schematize this kind of
data through simple graphs.  In order
to fit matching probabilities, in Section \ref{s:matchprobs} we
generalize the mixture model used by
\citet{Winkler88} and \citet{Jaro89}, and in Section \ref{s:EM} we
present details of the fitting of this model using the EM
algorithm (Dempster, Laird and Rubin, 1977).  In Section
\ref{s:assignment} we present an optimal method to decide the record
$K$-tuples membership to the subsets defined in Section
\ref{s:subsets}.  Section \ref{s:application} contains an application of the proposed methods to the integration of the three Colombian homicide record--systems and  Section \ref{s:simulation} describes simulation studies where we explore the performance of the method under different scenarios.

\section{COVERED SUBPOPULATIONS AND RECORD $K$-TUPLES}
\label{s:subsets}

We follow the exposition of \citet{FellegiSunter69} and
suppose some population is recorded by $K$ datafiles. Let
$A_1, A_2, \dots, A_K$ denote the $K$ overlapping subpopulations recorded in those $K$ datafiles.  Now, suppose that for each
datafile there exists one different record generating process
$\alpha_k$, which produces a set of records denoted
by
\[\alpha_k(A_k)=\{\alpha_k(a_k); a_k \in A_k\}, k=1,\dots,K\]
where the member $\alpha_k(a_k)$ represents a vector of information of the member $a_k \in A_k$.  This information could be subject to measurement error or
incomplete. Let us define the $K$-ary cartesian product
\begin{eqnarray*}
\bigotimes_{k=1}^K \alpha_k(A_k)
&=& \left\{\bigl(\alpha_1(a_1),\alpha_2(a_2),\dots,\alpha_K(a_K)\bigr); a_k \in A_k, k=1,\dots,K\right\}
\end{eqnarray*}
which is composed by all the possible record $K$-tuples in which the $k$th entry corresponds to the information recorded for some $a_k$ in the subpopulation $k$. Now we describe the possible matching patterns of the record $K$-tuples in terms of the members of the subpopulations $A_k$.  First, it is possible that a record $K$-tuple includes information on $K$ different individuals, i.e., for some $(\alpha_1(a_1),\alpha_2(a_2),\dots,\alpha_K(a_K)\bigr)$, $a_k\neq a_{k'}$, for all $k\neq k'$.  At the other extreme, if an individual appears in all $K$ datafiles, then in the record $K$-tuple $(\alpha_1(a_1),\alpha_2(a_2),\dots,\alpha_K(a_K)\bigr)$ actually $a_1 = a_2 = \dots = a_K$.  In general, we can classify the entries of each record $K$-tuple into subsets that record information on the same individual.

In order to establish this idea formally, let $\mathbb{P}_K$ denote the set of partitions of the set $\mathbb{N}_K=\{1,2,\dots,K\}$.  If we associate each number in $\mathbb{N}_K$ with an entry in a record $K$-tuple, then the matching pattern of each record $K$-tuple corresponds to a partition of $\mathbb{N}_K$, where the elements of the partition group the entries of the $K$-tuple that represent the same individual.  Now, let $S_{p}$ denote the set of record $K$-tuples corresponding to the matching pattern $p \in \mathbb{P}_K$.  It is clear that
\begin{equation}\label{eq:Kprod}
\bigotimes_{k=1}^K \alpha_k(A_k) = \bigcup_{p \in \mathbb{P}_K} S_{p}
\end{equation}
since each record $K$-tuple has a unique matching pattern.  The number of ways we can partition a set of $K$ elements into nonempty subsets is called the $K$th Bell number, denoted $B_K$, which can be found using the recurrence relation $B_K=\sum_{k=0}^{K-1}B_k\binom{K-1}{k}$, with $B_0=1$ by convention \citep[see][for further details]{Rota64}.  Thus, there are $B_K$ subsets $S_{p}$ of record $K$-tuples.

Let $n$ denote the cardinality of the set in equation (\ref{eq:Kprod}).  Also, for $j=1,\dots,n$,
let $r_j=\bigl(\alpha_1(a_1),\dots,\alpha_K(a_K)\bigr)$ for
some $a_k \in A_k, k=1,\dots,K$, be the $j$th record $K$-tuple of
the $K$-ary product in equation (\ref{eq:Kprod}).  When the datafiles do not contain common identifiers, we cannot identify the subset $S_{p}$ to which the record $K$-tuple $r_j$ belongs.  If the datafiles record the same $F$ fields of
information, however, we can obtain a comparison vector $\gamma^j$ for each $K$-tuple $r_j$.  We can use
this information to estimate the
probability that each record $K$-tuple belongs to each subset
$S_{p}$, given the comparison vector $\gamma^j$.  Multiple record linkage's goal is to classify all the record $K$-tuples in the appropriate subsets $S_p$.

\begin{figure*}[hb]
\centering
  \centerline{\includegraphics[trim = 1cm 15cm 1cm 8cm, width=0.8\linewidth]{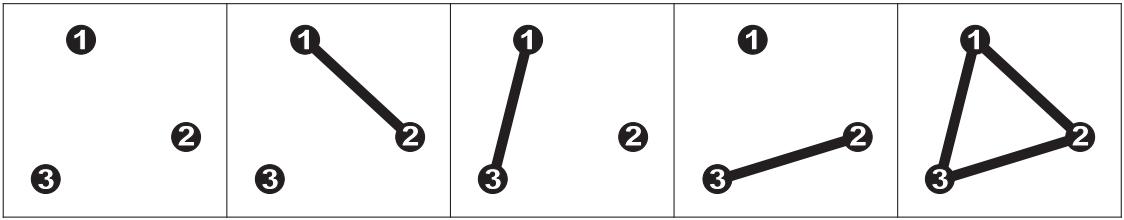}}
  \begin{minipage}[b]{0.85\textwidth}
  \caption{Undirected graphs giving $B_3=5$ possible patterns of agreement using three datafiles. The vertices appear connected if the value that each one represents agree, otherwise, the vertices appear unconnected.
  }
\label{f:Bell3}\end{minipage} 
\end{figure*}

\textit{Example.} If we have $K=3$ datafiles, for each triplet of records we have the matching patterns in Table \ref{t:partitions}, which can be represented using undirected graphs as in Figure \ref{f:Bell3}.  In this case, we also have $B_3=5$ and the cartesian product of the three datafiles can be written as
\begin{equation}
\bigotimes_{k=1}^3 \alpha_k(A_k) = S_{1/2/3} \cup S_{12/3} \cup S_{13/2} \cup S_{1/23} \cup S_{123}.
\end{equation}

\begin{table}[hb]
 \vspace*{-6pt}
  \centering
  \begin{minipage}[b]{0.5\textwidth}
 \def\~{\hphantom{0}}
  \centering
  \caption{Each matching pattern of a record triplet can be associated with a partition of the set $\{1,2,3\}$.} \label{t:partitions}
\vspace{.1cm}
\begin{tabular}{lll}
\hline\hline\\[-8pt]
Notation & $\mathbb{P}_3$ & $(\alpha_1(a_1),\alpha_2(a_2),\alpha_3(a_3)\bigr)$\\
\\[-6pt]\hline
1/2/3 & $\{\{1\},\{2\},\{3\}\}$ & $a_1 \neq a_2 \neq a_3 \neq a_1$\\
12/3 & $\{\{1,2\},\{3\}\}$ & $a_1 = a_2$; $a_3\neq a_1,a_2$\\
13/2 & $\{\{1,3\},\{2\}\}$ & $a_1 = a_3$; $a_2\neq a_1,a_3$\\
1/23 & $\{\{1\},\{2,3\}\}$ & $a_2 = a_3$; $a_1\neq a_2,a_3$\\
123 & $\{\{1,2,3\}\}$ & $a_1 = a_2 = a_3$\\
\hline
\end{tabular}

\end{minipage}
\end{table}

\subsection{Blocking}\label{ss:blocking}
Note that the dimension of the $K$-ary product grows exponentially as a function of $K$.  Thus, considering the complete set of record $K$-tuples is highly inefficient in most applications.  A common way to deal with this problem in bipartite record linkage is to partition each datafile into a common set of blocks, thereby eliminating the need to match records in different blocks.  The idea is that reliable categorical fields such as zip code or gender may be used to quickly label some of the non-links.  For example, if we are matching datafiles with geographic information, we could assign those records that differ in zip code (or a similar field) as non-links.  See \citet{HerzogScheurenWinkler07, HerzogScheurenWinkler10} and \citet{Christen11} for a discussion of blocking.

In multiple record linkage we can apply the same idea to assign non--links between pairs of records within every record $K$-tuple.  If a certain blocking variable assigns a non--link between records $k$ and $k'$ in the record $K$-tuple $r_j$, this implies that $r_j$ cannot be assigned to subsets $S_p$ where the pattern of agreement $p$ involves a link between files $k$ and $k'$.  Consequently, the record linkage process has to decide among the remaining possibilities.  If a non--link is assigned to every pair of records within a record $K$-tuple, then this $K$-tuple can be assigned directly to the subset $S_{1/2/\dots/K}$ (see notation in Table \ref{t:partitions}).  In practice this last step tremendously reduces the number of $K$-tuples to be classified.

Using the natural partial order in $\mathbb{P}_K$ we provide a way to determine the subsets to which a record $K$-tuple can be assigned after blocking.  We say that $p'\preccurlyeq p$ if $p'$ is a partition finer than or equal to $p$.  Note that the blocking process provides a maximal pattern of agreement $p_b$ for each record $K$-tuple $r_j$.  Thus, the subsets to which $r_j$ can be potentially assigned are those $S_p$ such that $p\preccurlyeq p_b$.

\begin{figure*}[hb]
\centering
  \centerline{\includegraphics[width=0.45\linewidth]{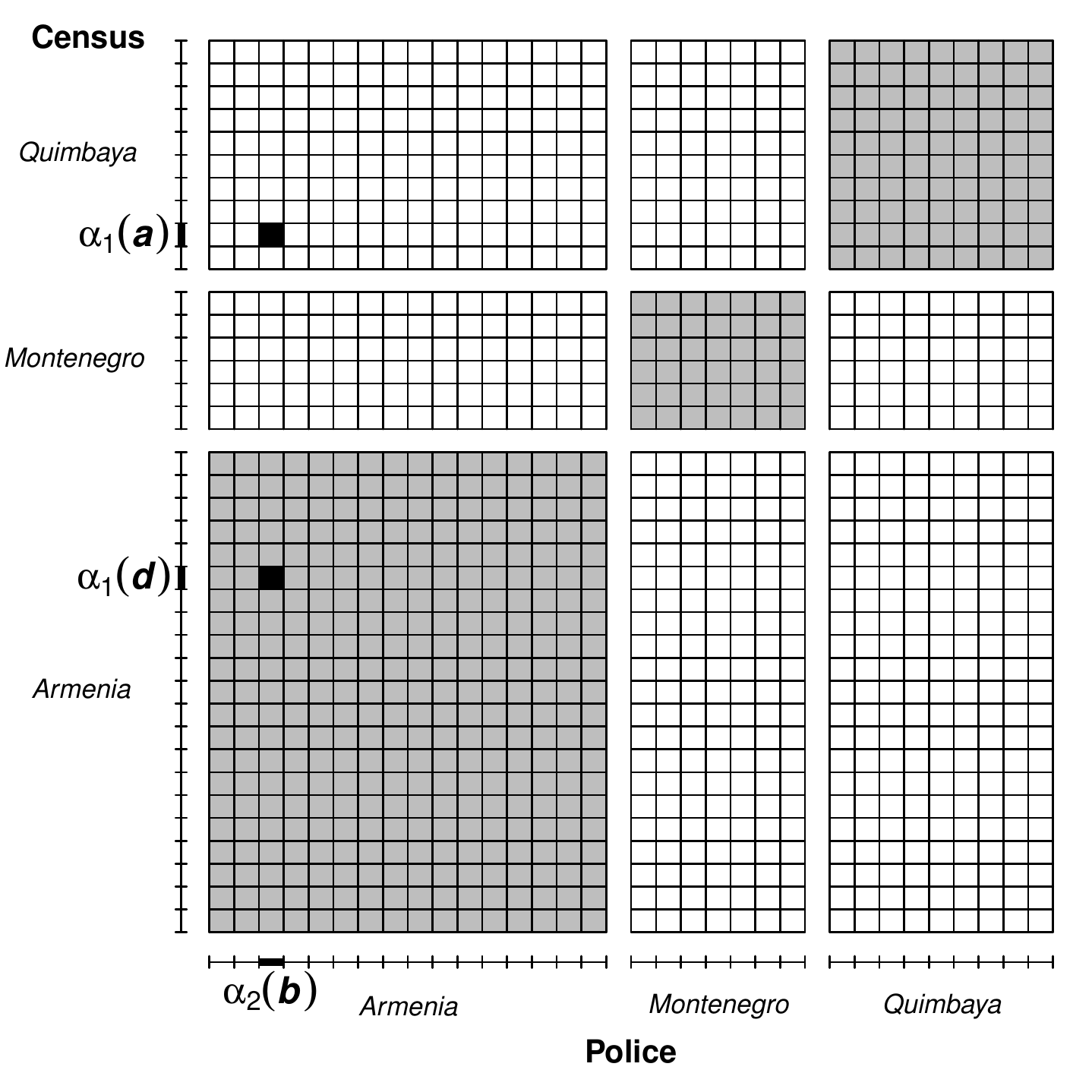}\hspace{.4cm}\includegraphics[width=0.45\linewidth]{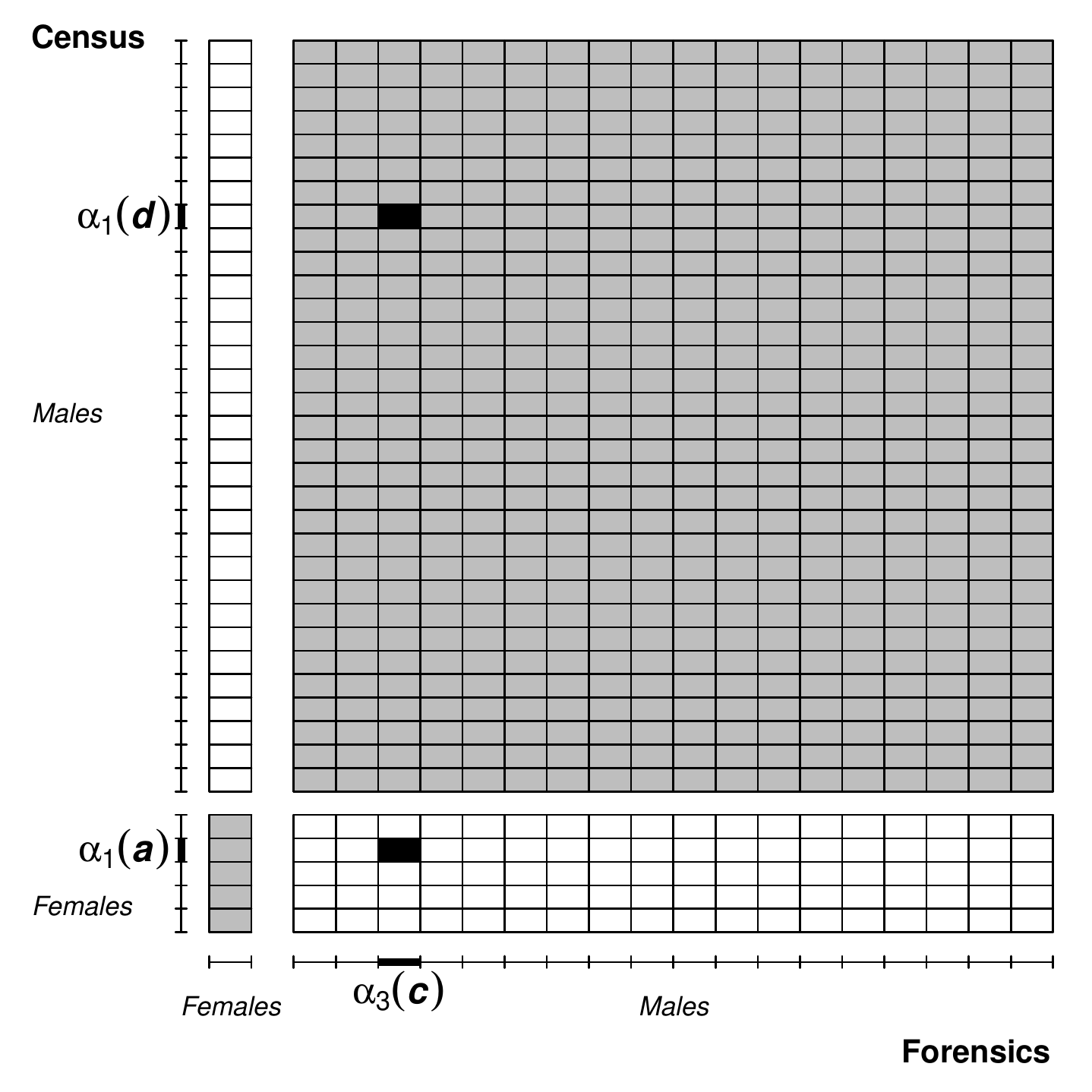}}
  \begin{minipage}[b]{0.85\textwidth}
  \caption{Cartesian products of Census and Police homicide data after blocking by town (left), and Census and Forensics homicide data after blocking by gender (right) for three towns in Colombia.  Only elements in gray blocks are potentially linked.  Black elements are discussed in the example of Section \ref{ss:blocking}.}
\label{f:blocking}\end{minipage} 
\end{figure*}

\textit{Example.}  In Figure \ref{f:blocking} we present the cartesian product of two pairs of files after blocking.  We illustrate using homicide data from the Armenia, Montenegro, and Quimbaya towns in the Colombian province of Quindio.  In this example, only the gray elements of the cartesian product become part of the record linkage process, whereas the white elements become a priori non--matches.  The left--hand side of Figure \ref{f:blocking} represents the cartesian product of two Census and Police data subsets after blocking by town.  The right--hand side represents the cartesian product of the same Census data subset and a Forensics data subset after blocking by gender.  Note that in this example we assign the pair $(\alpha_1(a),\alpha_2(b))$ as a non--link since these two records refer to homicides in different towns.  We also assign the pair $(\alpha_1(a),\alpha_3(c))$ as a non--link since these two records refer to different genders.  Assuming that there are no non--link blocking assignments for $(\alpha_2(b),\alpha_3(c))$, the multiple record linkage decision process has to classify the triplet $(\alpha_1(a),\alpha_2(b),\alpha_3(c))$ as either belonging to $S_{1/2/3}$ or $S_{1/23}$.  On the other hand, the two blocking processes illustrated in Figure \ref{f:blocking} have no direct implications on the possible resolution of $(\alpha_1(d),\alpha_2(b),\alpha_3(c))$.

\section{COMPARISON DATA}
\label{s:data}

In order to obtain appropriate data to model the probability that a certain record $K$-tuple belongs to some subset $S_{p}$, let us determine the matching pattern for each common field of recorded information.  If for a certain record $K$-tuple we search for agreement among the information recorded for a certain field, we can associate each entry of the record $K$-tuple with a number in $\{1,2,\dots,K\}$ and a certain partition of this set would describe the matching pattern of the record $K$-tuple for the field in consideration, grouping in the same element of the partition all the $K$-tuple entries that agree in the field being compared (similar to Section \ref{s:subsets}).  An alternative way to explain this idea is as follows. For some record $K$-tuple, let us compare the information of the records from the datafiles $k$, $k'$, and $k''$ for a certain common field.  Due to transitivity of agreement, if records $k$ and $k'$ agree and $k'$ and $k''$ agree, then $k$ and $k''$ agree necessarily.  Thus, since agreement is an equivalence relation, each matching pattern for each field for each record $K$-tuple is a partition of $K$ points, because for any equivalence relation on a set, the set of its equivalence classes (sets of records agreeing) is a partition of the set.

Now, let $\gamma^{j_f}_{p}=1$ if the record $K$-tuple $r_j$ has the matching pattern $p$ in the field $f$. Then, for each field $f=1,\dots,F$, of each record $K$-tuple $r_j$, we obtain a vector $\gamma^{j_f}=(\gamma^{j_f}_{1/2/\dots/K},\dots,\gamma^{j_f}_{p},\dots,\gamma^{j_f}_{12\dots K})$, where only one entry is equal to one and the rest are equal to zero. Note the length of the vector $\gamma^{j_f}$ is $B_K$, since this is the number of patterns of agreement for each field.  Finally, the comparison data for $r_j$ contains the comparison vectors for all the $F$ fields, and can be written as $\gamma^j=(\gamma^{j_1},\dots,\gamma^{j_f},\dots,\gamma^{j_F})$, which takes values over $(B_K)^F$ possible matching patterns.

Similarly as in Section \ref{s:subsets}, we can represent the patterns of agreement presented in this section by unions of complete undirected graphs \citep[see][p. 448]{Rosen06} as in Figure \ref{f:Bell3}. In those graphs, each vertex represents the value of certain field in certain record that belongs to certain datafile $k=1,\dots,K$. The
vertices $k'$ and $k$ appear connected if the values that they represent agree, otherwise, the vertices appear disconnected.

\textit{Example.}  Let us expose
how the comparison data work when we need to link three datafiles.  In this case, we can represent the patterns of
agreement as five unions of complete undirected
graphs, as presented in Figure \ref{f:Bell3}. For $K=3$, $\gamma^{j_{f}}=$ $(\gamma^{j_{f}}_{1/2/3},\gamma^{j_{f}}_{12/3},\gamma^{j_{f}}_{13/2},\gamma^{j_{f}}_{1/23},\gamma^{j_{f}}_{123})$ represents the
comparison data for the field $f$ (say age, ethnicity, etc.) of the record triplet $r_j$, and the length of the full comparison data for each record triplet is $5F$, if the datafiles have $F$ common fields.

\section{MODEL FOR MATCHING PROBABILITIES}
\label{s:matchprobs}

The probabilities $P(S_{p}| \gamma^j )\doteq P(r_j\in S_{p}| \gamma^j )$, $p \in \mathbb{P}_K$, can be found using $P(\gamma^j|S_{p})\doteq P(\gamma^j | r_j\in S_{p})$ and $P(S_{p})\doteq P(r_j\in S_{p})$, as $P(S_{p} | \gamma^j )=P(\gamma^j|S_{p})P(S_{p})/P(\gamma^j)$,
where
\begin{equation*}\label{eq:pGamma}
P(\gamma^j) = \sum_{p \in \mathbb{P}_K}P(\gamma^j|S_{p})P(S_{p}).
\end{equation*}
Let $g^j=(g^j_{1/2/\dots /K},\dots,g^j_{12\dots K})$ be the vector that indicates the subset $S_{p}$ that contains the record $K$-tuple $r_j$, such
that $g^j_{p}= 1 \text{ if } r_j\in S_{p}$ and  $g^j_{p}= 0$ otherwise.  Thus, it is clear that $\sum_{\mathbb{P}_K} g^j_{p}=1$. Now, let $x^j=(g^j,\gamma^j)$ be the (partially observed) complete data vector for $r_j$.  Note that after blocking, some entries of $g^j$ are fixed as zeroes for some record $K$-tuples.

\citet{Winkler88}, \citet{Jaro89}, and  \citet{LarsenRubin01} proposed to model the corresponding complete data $x^j$ for bipartite record linkage, where $g^j$ is taken as a latent variable.  For multiple record linkage, the model for $x^j$ is stated as
\begin{equation*}
P(x^j|\Phi)
= \prod_{p \in \mathbb{P}_K}\Bigl[P(\gamma^j| S_{p})P( S_{p})\Bigr]^{g^j_{p}}.
\end{equation*}
Under the conditional independence assumption of the comparison data fields, we obtain
\begin{equation}\label{eq:indep1}
P(\gamma^j|S_{p})=\prod_{f=1}^F P(\gamma^{j_f}| S_{p}).
\end{equation}
Each $\gamma^{j_f}$ represents the matching pattern of $r_j$ in the field $f$, which corresponds to categorical information that can be
modeled by using a categorical distribution (or multinomial with just one trial) as
\begin{equation}\label{eq:multinom}
P(\gamma^{j_f}| S_{p}) = \prod_{p' \in \mathbb{P}_K} (\pi ^f_{p'|p})^{\gamma^{j_f}_{p'}}
\end{equation}
where $\pi ^f_{p'|p}\doteq P(\gamma^{j_f}_{p'}=1| S_{p})$, and $p'$ is just another indicator of the patterns of agreement in $\mathbb{P}_K$.  Defining $s_{p} \doteq P(S_{p})$, under independence of the complete data, the complete log--likelihood for the sample $\textbf{x}=\{x^j; j=1,\dots,n\}$ is obtained as
\begin{eqnarray*}
L &=&\sum_{j=1}^n\sum_{p\in \mathbb{P}_K}g^j_{p}\Bigl[\log s_{p} +\sum_{f=1}^F \sum_{p'\in \mathbb{P}_K} \gamma^{j_f}_{p'}\log \pi ^f_{p'|p}\Bigr].
\end{eqnarray*}
The set of parameters in the log--likelihood above is $\Phi=(\mathbf{s},\Pi)$, where $\mathbf{s}$ is a vector of length $B_K$ given by
$\mathbf{s}=(s_{1/2/\dots/K},\dots,s_{12\dots K})$ and $\Pi$ can be arranged in a set of $F$ matrices of size
$B_K\times B_K$, each one given by
\begin{eqnarray*}
\Pi^f&=&\begin{pmatrix}
\pi^{f}_{1/2/\dots /K|1/2/\dots /K}&\dots&\pi^{f}_{1/2/\dots /K|p}&\dots&\pi^{f}_{1/2/\dots /K|12\dots K}\\
\vdots&\ddots&\vdots&\ddots&\vdots\\
\pi^{f}_{p'|1/2/\dots /K}&\dots&\pi^{f}_{p'|p}&\dots&\pi^{f}_{p'|12\dots K}\\
\vdots&\ddots&\vdots&\ddots&\vdots\\
\pi^{f}_{12\dots K|1/2/\dots /K}&\dots&\pi^{f}_{12\dots K|p}&\dots&\pi^{f}_{12\dots K|12\dots K}\\
\end{pmatrix}
\end{eqnarray*}
for $f=1,\dots,F$.  Hence, the length of $\Phi$ is $B_K(B_KF+1)$.  In order to estimate these probabilities, since the $g^j$ vectors are only partially observed, the estimation is made via maximum likelihood using the EM algorithm \citep{DempsterLairdRubin77}.  The model presented in this section generalizes the one used by \citet{Winkler88} and \citet{Jaro89}, and uses the strong assumption that the comparison data fields are conditionally independent given the $K$-tuples' membership to the subsets $S_p$.  In Section \ref{s:application} we show that this baseline model produces good results for the Colombian homicide data, but the modeling of the fields' dependencies may be a key factor in obtaining good linkage results in other contexts \citep[see][]{LarsenRubin01}.  This is part of our ongoing work.

\textit{Example.} For the particular case where $K=3$, the length of $\Phi$ is $5+25F$, which is given by
$\mathbf{s}=( s_{1/2/3},s_{12/3},s_{13/2},s_{1/23},s_{123})$ and $\Pi$, which is composed by $F$ matrices of size $5\times 5$, as
\begin{eqnarray*}
\Pi^f&=&\begin{pmatrix}
\pi^{f}_{1/2/3|1/2/3}&\pi^{f}_{1/2/3|12/3}&\pi^{f}_{1/2/3|13/2}&\pi^{f}_{1/2/3|1/23}&\pi^{f}_{1/2/3|123}\\
\pi^{f}_{12/3|1/2/3}&\pi^{f}_{12/3|12/3}&\pi^{f}_{12/3|13/2}&\pi^{f}_{12/3|1/23}&\pi^{f}_{12/3|123}\\
\pi^{f}_{13/2|1/2/3}&\pi^{f}_{13/2|12/3}&\pi^{f}_{13/2|13/2}&\pi^{f}_{13/2|1/23}&\pi^{f}_{13/2|123}\\
\pi^{f}_{1/23|1/2/3}&\pi^{f}_{1/23|12/3}&\pi^{f}_{1/23|13/2}&\pi^{f}_{1/23|1/23}&\pi^{f}_{1/23|123}\\
\pi^{f}_{123|1/2/3}&\pi^{f}_{123|12/3}&\pi^{f}_{123|13/2}&\pi^{f}_{123|1/23}&\pi^{f}_{123|123}\\
\end{pmatrix}.
\end{eqnarray*}

\section{EM ESTIMATION}
\label{s:EM}

The EM algorithm can be used to fit the parameters of a mixture model via maximum likelihood estimation \citep[see][p.
47]{DempsterLairdRubin77,McLachlanPeel00} and has been applied to record linkage problems \citep[e.g.,][]{Winkler88,Jaro89,LarsenRubin01}.
Following the model presented in Section \ref{s:matchprobs}, let us find the equations of an EM algorithm to estimate $\Phi$. Firstly, for the
\textbf{E}xpectation step, let us find the conditional distribution of $g^j$
\begin{eqnarray*}
P(g^j | \gamma^j)&=&\frac{P(x^j)}{P(\gamma^j)}\nonumber \\
&=& \prod_{p\in \mathbb{P}_K}\left[\frac{P(\gamma^j| S_{p})P( S_{p})}{P(\gamma^j)}\right]^{g^j_{p}}\\
&=& \prod_{p\in \mathbb{P}_K}\left[P( S_{p}|\gamma^j)\right]^{g^j_{p}}
\end{eqnarray*}
i.e., $g^j|\gamma^j\sim Multinomial\bigl(1, P|\gamma^j\bigr)$, where $P|\gamma^j=\left(P( S_{1/2/\dots /K}| \gamma^j), \dots, P( S_{12\dots K}| \gamma^j )\right)$.

Thus, using the estimation $\hat \Phi$ from a previous M step of the algorithm, for the E step, the expectation of the unknown part of $g^j$ is composed by
\begin{equation}\label{eq:ghat}
\hat P( S_{p} | \gamma^j )=\frac{ \hat{s}_{p} \prod_{f=1}^F\prod_{p'\in \mathbb{P}_K} (\hat \pi ^f_{p'|p})^{\gamma^{j_f}_{p'}}  }{\hat P(\gamma^j)}
\end{equation}
for $p\preccurlyeq p_{b_j}$, where $p_{b_j}$ represents the blocking pattern for $r_j$. The term $\hat P(\gamma^j)$ above is given by
\begin{equation*}
\hat P(\gamma^j)=\sum_{p\preccurlyeq p_{b_j}} \hat{s}_{p} \prod_{f=1}^F\prod_{p'\in \mathbb{P}_K} (\hat \pi ^f_{p'|p})^{\gamma^{j_f}_{p'}}.
\end{equation*}

Let $\tilde g^j$ be equal to $g^j$ for the entries that are known to be zeroes, and let the remaining entries of $\tilde g^j$ be filled with the values given in equation (\ref{eq:ghat}).  For the \textbf{M}aximization step, we replace $g^j$ with $\tilde g^j$ in the log--likelihood $L$ and estimate $\Phi$ via maximum likelihood.  We obtain for $\hat\Pi$
\begin{equation*}
\hat \pi ^f_{p'|p}
=\frac{\sum_{j=1}^{(B_K)^F}n_{\gamma^j}\gamma^{j_f}_{p'}\tilde g^j_p}{\sum_{j=1}^{(B_K)^F}n_{\gamma^j}\tilde g^j_p},
\end{equation*}
and for $\hat{\mathbf{s}}$ we obtain
\begin{equation*}
\hat{s}_{p}
=\frac{\sum_{j=1}^{(B_K)^F} n_{\gamma^j} \tilde g^j_p}{n}
\end{equation*}
where $n_{\gamma^j}$ represents the frequency counts of each pattern
$\gamma^j$, as in \citet{Jaro89}. Note that in this case we have $(B_K)^F$ different patterns of $\gamma^j$.  As usual, the algorithm stops when
the values of $\hat \Phi$ converge, which can be assessed measuring the distance between $\hat \Phi$ in two consecutive iterations.  In order to start this algorithm, we choose initial values taking
into account the fact that some probabilities must be greater than
others.

\subsection{Starting Values}

Note that the parameters in each $\Pi^f$ should hold certain restrictions.  In record linkage these constraints are taken into account in order to start the EM algorithm \citep{Winkler93,LahiriLarsen05}.  For instance, it is clear that $\pi ^f_{12\dots K|12\dots K}$ should
be greater than $\pi ^f_{1/2/\dots /K|12\dots K}$, i.e., given that in a record $K$-tuple all the entries refer to the same individual, the probability that their information agree should be larger than the probability that all their information disagree.  However, note that $\pi ^f_{1/2/\dots /K|1/2/\dots /K}$ should not necessarily be greater than $\pi ^f_{12\dots K|1/2/\dots /K}$, i.e., for a record $K$-tuple in which all the entries refer to different individuals, the probability that all their information disagree is not necessarily larger than the probability that all their information agree (this is the case for a field with a very common value).

Thus, given the high number of parameters it is not easy to determine which constraints should be taken into account.  In order to determine the set of constraints to start the algorithm, we present a method that uses the natural partial order in $\mathbb{P}_K$.  Remember that we say $p'\preccurlyeq p$ if $p'$ is a partition finer than or equal to $p$.  In order to determine if $\pi^f_{p'|p}$ should be greater or lower than $\pi^f_{p''|p}$ for $p,p',p''\in \mathbb{P}_K$, we fix the partition $p$ and for all partitions $p', p''$, such that $p''\preccurlyeq p'\preccurlyeq p$ we set $\pi^f_{p''|p}\leq \pi^f_{p'|p}$.  In any other case we do not have a criterion to order $\pi^f_{p'|p}$ with respect to $\pi^f_{p''|p}$.

Note that this procedure can be visualized using a directed graph in the following way:
\begin{itemize}
\item[1.] Construct the Hasse diagram of the partitions $p'\in\mathbb{P}_K$ writing in each node $\pi^f_{p'|p}$ where $p$ is a generic partition.
\item[2.] Assign a specific partition to the generic $p$.
\item[3.] Search for the node where $p'=p$.
\item[4.] For all the branches under this node, set an inequality $\geq$ between each ``father'' node and each ``son'' node.
\item[5.] Repeat steps 2 -- 4 until exhausting the possible partitions.
\end{itemize}

We can use similar ideas to identify the constraints for $s_{1/2/\dots/K},\dots,s_{12\dots K}$.  We simply have that $s_{p'} \geq s_{p}$ whenever $p'\preccurlyeq p$.  Naturally, the set of inequalities among the probabilities $s_{p}$ can also be represented in a Hasse diagram.  Furthermore, if the datafiles being linked have no duplicates, the size of the complete links set $S_{12\dots K}$ should be smaller than or equal to the smaller datafile size, from which is reasonable to take starting values for $s_{12\dots K}$ smaller than $\min \{ m_k; k=1,\dots,K\}/n$, where $m_k$ represents the number of records in datafile $k$.  In general we can determine the maximum size of any set $S_p$ if we assume no duplicates into each datafile.  Denote $q_p$ as a generic element of the partition $p\in\mathbb{P}_K$, i.e., $q_p$ is a subset of $\mathbb{N}_K$.  Thus, the maximum size of $S_p$ is $\prod_{q_p\in p} \min \{ m_k; k\in q_p\}$, from which is reasonable to start the algorithm taking values lower than $\prod_{q_p\in p} \min \{ m_k; k\in q_p\}/n$ for a generic $s_p$.  The starting value for $s_{1/2/\dots/K}$ is determined as one minus the other $s_p$.  Notice that  since duplicates are rather common in practice, the above values are merely a guide to start the EM algorithm.  Finally, since latent class models have multiple solutions corresponding to local maxima of the marginal likelihood, in practice we would take different starting values holding the above constraints, and we would choose the parameters with the maximum marginal likelihood for the observed data $\gamma^j$ \citep[e.g., see][]{McLachlanPeel00}.

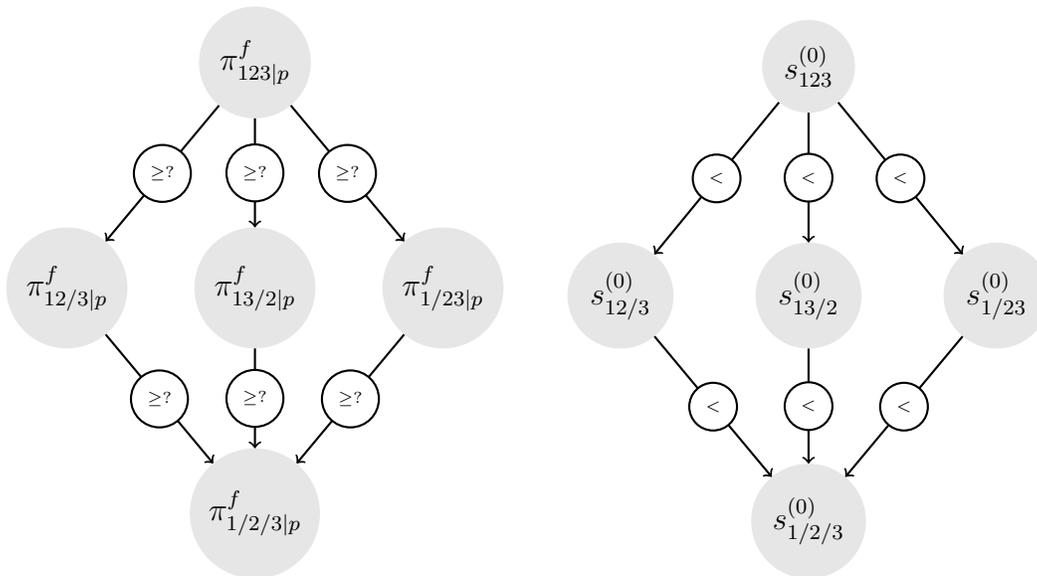
\begin{figure*}[hb]
\centering
\begin{tikzpicture}
  \matrix (galois)
    [matrix of nodes,%
     nodes in empty cells,
     nodes={outer sep=0pt,circle,minimum size=0pt,fill=gray!20},
     column sep={2.5cm,between origins},
     row sep={3cm,between origins}]
  {
   |[draw=none,fill=none]| & $\pi^f_{123|p}$ & |[draw=none,fill=none]|\\
    $\pi^f_{12/3|p}$ &  $\pi^f_{13/2|p}$ &  $\pi^f_{1/23|p}$\\
   |[draw=none,fill=none]| &  $\pi^f_{1/2/3|p}$ & |[draw=none,fill=none]|\\
  };
  \foreach \a in {1,3}
    \draw[->, thick] (galois-1-2) -- (galois-2-\a) node [circle,midway,fill=white,draw] {\tiny{$\geq ?$}};
  \foreach \a in {1,3}
    \draw[->, thick] (galois-2-\a) -- (galois-3-2) node [circle,midway,fill=white,draw] {\tiny{$\geq ?$}};
    \draw[->, thick] (galois-1-2) -- (galois-2-2) node [circle,midway,fill=white,draw] {\tiny{$\geq ?$}};
    \draw[->, thick] (galois-2-2) -- (galois-3-2) node [circle,midway,fill=white,draw] {\tiny{$\geq ?$}};
\end{tikzpicture}  \hspace{.3cm} \begin{tikzpicture}
  \matrix (galois)
    [matrix of nodes,%
     nodes in empty cells,
     nodes={outer sep=0pt,circle,minimum size=0pt,fill=gray!20},
     column sep={2.5cm,between origins},
     row sep={3cm,between origins}]
  {
   |[draw=none,fill=none]| & $s_{123}^{(0)}$ & |[draw=none,fill=none]|\\
    $s_{12/3}^{(0)}$ &  $s_{13/2}^{(0)}$ &  $s_{1/23}^{(0)}$\\
   |[draw=none,fill=none]| &  $s_{1/2/3}^{(0)}$ & |[draw=none,fill=none]|\\
  };
  \foreach \a in {1,3}
    \draw[->, thick] (galois-1-2) -- (galois-2-\a) node [circle,midway,fill=white,draw] {\tiny{$<$}};
  \foreach \a in {1,3}
    \draw[->, thick] (galois-2-\a) -- (galois-3-2) node [circle,midway,fill=white,draw] {\tiny{$<$}};
    \draw[->, thick] (galois-1-2) -- (galois-2-2) node [circle,midway,fill=white,draw] {\tiny{$<$}};
    \draw[->, thick] (galois-2-2) -- (galois-3-2) node [circle,midway,fill=white,draw] {\tiny{$<$}};
\end{tikzpicture}

  \begin{minipage}[b]{0.7\textwidth}
  \caption{Hasse diagram to determine the set of inequalities between probabilities $\pi^f_{p'|p}$ and $s_p^{(0)}$.  The possible inequalities are established from sources to targets in the arrows, e.g., $s_{123}^{(0)}<s_{12/3}^{(0)}$.
  }
\label{f:Hasse}\end{minipage} \vspace{.5cm}
\end{figure*}

\textit{Example.}  We illustrate this procedure for $K=3$ using the left hand--side of Figure \ref{f:Hasse}.  Go to the left panel of Figure \ref{f:Hasse} and replace $p$ with $123$.  Since $\pi^f_{123|123}$ is in the top of the graph, we take the set of constraints
$\pi^f_{123|123}\geq\pi^f_{12/3|123}\geq\pi^f_{1/2/3|123}; \ \pi^f_{123|123}\geq\pi^f_{13/2|123}\geq\pi^f_{1/2/3|123}; \ \pi^f_{123|123}\geq\pi^f_{1/23|123}\geq\pi^f_{1/2/3|123}$, which correspond to the three different branches under $\pi^f_{123|123}$.  Now, replace $p$ with $12/3$.  Since in this case the node $\pi^f_{12/3|12/3}$ has only one descendent, we only get the constraint $\pi^f_{12/3|12/3}\geq\pi^f_{1/2/3|12/3}$.  This step is similar for $13/2$ and $1/23$.  Finally, if we replace $p$ with $1/2/3$ we can see that the node $\pi^f_{1/2/3|1/2/3}$ does not have descendants, so we do not set constraints for the probabilities $\pi^f_{p'|1/2/3}$, $p' \in \mathbb{P}_3$.

For $K=3$, the right hand side of Figure \ref{f:Hasse} represents the set of inequalities for the starting values $s_{p}^{(0)}$. We obtain for instance $s_{1/2/3}^{(0)}>s_{1/23}^{(0)}>s_{123}^{(0)}$.  Also, for this particular  case we take $s_{123}^{(0)} <\min\{m_1,m_2,m_3\}/n$, \ $s_{1/23}^{(0)} <m_1\min\{m_2,m_3\}/n$, and similar inequalities for $s_{13/2}^{(0)}$ and $s_{12/3}^{(0)}$, whereas $s_{1/2/3}^{(0)}=1-s_{1/23}^{(0)}-s_{13/2}^{(0)}-s_{12/3}^{(0)}-s_{123}^{(0)}$.

\section{LINKAGE ASSIGNMENT: GENERALIZED \\ FELLEGI--SUNTER DECISION RULE}
\label{s:assignment}

The goal of multiple record linkage is to classify each record
$K$-tuple to the appropriate subset $S_{p}$. For bipartite record linkage, \citet{FellegiSunter69}
proposed the computation of likelihood ratios as weights for the
assignment of record pairs as matched or unmatched pairs.  Their
procedure is equivalent to test the hypothesis that each record pair
belongs to the subset of unmatched record pairs, against the
hypothesis that it belongs to the subset of matched pairs, and vice
versa.

\subsection{Likelihood Ratios and Weights}
In multiple record linkage, there are several subsets of records denoting all the possibilities of matching between records from different datafiles.
Following Fellegi and Sunter's idea, for each record $K$-tuple and
for each subset, we propose to compute weights following a
hypothesis test, where the null hypothesis is the record $K$-tuple
membership to a certain subset, i.e., $r_j\in S_p$, against the hypothesis that this
record $K$-tuple does not belong to the subset, i.e., $r_j\in S^c_p$, where the superscript $c$ denotes the complement of the set.  By using a
log--likelihood ratio we obtain
\begin{equation*}
w_{p}^j=\log\frac{ P(\gamma^j|S_{p})}{ P(\gamma^j|S^c_{p})}.
\end{equation*}
The informal idea of the use of the weights $w_{p}^j$ is that we would order the record $K$-tuples according to their respective weights and we would assign $K$-tuples with large $w_{p}^j$ to the subset $S_p$.  However, the ordering obtained from $w_{p}^j$ can be obtained in a simpler way, regardless of the model for $P(\gamma^j|S_{p})$.\\

\textit{Proposition 1.}  The ordering obtained from $w_{p}^j$, $\text{logit}[P(S_{p}|\gamma^j)]$ and $P(S_{p}|\gamma^j)$ is the same.\\

Thus, for ordering and decision purposes we can simply use $P(S_{p}|\gamma^j)$ (see proofs in the Appendix \ref{ap:optimrule}).  We still need to determine, however, the cutoffs from which we declare record $K$-tuples' memberships.

\subsection{Cutoff Values}\label{ss:cutoff}

In bipartite record linkage, in order to declare a record pair as matched or unmatched, the
Fellegi--Sunter method orders the possible
values of $\gamma^j$ by their weights in non--increasing order,
determines two cutoff values of the weights, and, according to them, declares matches and non--matches.  For multiple record
linkage, we extend this procedure and prove its optimality.\\

\textit{Theorem 1.}  The decision procedure described below maximizes the probability of assigning each record $K$-tuple to the right subset, subject to a set of admissible error levels $\mu_p$.
\begin{itemize}
\item[1.] Each record $K$-tuple is \emph{potentially} declared to belong to the subset $S_p$ if and only if $p$ is the pattern for which $P(S_{p}|\gamma^j)$ is maximum among all possible patterns in $\mathbb{P}_K$.  Thus, the set of record $K$-tuples is partitioned into $B_K$ subsets, and for each record $K$-tuple in one of these partitions we consider only two possibilities, whether to declare it to belong to the subset $S_p$ or to keep it undeclared.
\item[2.] For the record $K$-tuples in each partition, we order the possible values of
$\gamma^j$ by their weights (or equivalently by $P(S_{p}|\gamma^j)$) in non--increasing order indexing by the
subscript $(j)_p$.
\item[3.] We find one value $(j')_p$ for each set of
weights related to each subset, in order to determine the record
$K$-tuple memberships. The value $(j')_p$ is found such that
\[
\mu_p=\sum_{(j)_p=1}^{(j')_p-1} P(\gamma^{(j)_p}| S^c_{p})
\]
where $\mu_p=P(\text{assign } r_j \text{ the membership of } S_{p}|r_j\in S^c_{p})$ is an admissible error level. Each
$P(\gamma^{(j)}|S^c_{p})$ can be computed as
\begin{eqnarray*}
P(\gamma^{(j)_p}| S^c_{p})&=&\frac{\sum_{p' \in \mathbb{P}_K, p'\neq p} P(\gamma^{(j)_p}| S_{p'}){s}_{p'}}{1- {s}_{p}}.
\end{eqnarray*}
\item[4.] Finally, for those record $K$-tuples with configurations of $\gamma^{(j)_p}$, $(j)_p=1,\dots,(j')_p-1$,
we decide that they belong to the subset $S_p$.  For those record $K$-tuples with configurations $\gamma^{(j)_p}$ with $(j)_p\geq(j')_p$, we keep them undeclared.
\end{itemize}

In the Appendix we show that the above decision rule is optimal under the availability of the true matching probabilities.  We show that this decision rule minimizes the probability of assigning each record $K$-tuple to the wrong subset $S_p$ or keeping it undeclared, subject to a set of admissible error levels $\mu_p$, or namely, it maximizes the probability of assigning each record $K$-tuple to the right subset, subject to a set of admissible error levels $\mu_p$.  The Fellegi--Sunter decision rule for bipartite record linkage can be obtained as a corollary of Theorem 1.  In practice the optimality of this decision rule depends on the quality of the estimation of the matching probabilities.  \citet{BelinRubin95} and \citet{LarsenRubin01} provide evidence that nominal and actual error levels disagree in different applications.  \citet{BelinRubin95} proposed a method to calibrate error rates as a function of cutoff values for bipartite record linkage.  This is an important problem that we expect to address in our ongoing work for the multiple record linkage context.

\section{LINKING HOMICIDE RECORD--SYSTEMS IN \\ COLOMBIA}\label{s:application}

The Colombian homicide data described in Section \ref{ss:recordsystems} was provided by the Conflict Analysis Resource Center (CERAC) where a linkage by hand was performed for a subset of the data, corresponding to the province of Quindio for the last three months of 2004.  In this section we present an application to the integration of these three datafiles.  In this period, 67, 62, and 33 homicides were recorded by the Census Bureau, the National Police, and the Forensics Institute, respectively.  The common fields of these three datafiles are town and date of the homicide, gender, and age of the victim.

An outline of the implementation of the method is as follows:
\begin{itemize}
\item[1.] Find the set of record triplets that are suitable for classification into the different matching patterns.  This set is obtained after blocking.
\item[2.] Compute the comparison data according to the possible patterns of agreement for all the triplets to be classified and for every common field.
\item[3.] Train the mixture model of the distribution of the comparison data.
\item[4.] Divide the set of triplets according to the subsets $S_p$ for which $\hat P(S_p|\gamma^j)$ is maximum.
\item[5.] Within each subset, sort the triplets by $\hat P(S_p|\gamma^j)$ and use an admissible error level to either declare the triplets as belonging to the subset $S_p$ or keep them undeclared.
\end{itemize}

In order to implement the method, we used town of the homicide and gender of the victim for blocking.  We assigned the membership to the subset $S_{1/2/3}$ to the triplets with blocking pattern $1/2/3$.  We used the proposed method to classify the remaining triplets.   In order to use date of the homicide and age of the victim, we explored several options, but we only report the results of using three of them (Table \ref{t:agedateoptions}).  The first option only includes exact comparison data for both variables.  The second option constructs three categorical variables from each variable age and date, and creates comparison data using these new categorical variables.  These variables are constructed in the following fashion: The categories of the variable AgeA are 0--2, 3--5, and so on; the categories of the variable AgeB are 0, 1--3, 4--6, and so on; and finally, the categories of the variable AgeC are 0--1, 2--4, and so on.  A similar procedure is used for date of the homicide, starting from the first day of the period of the data.  The third approach uses the previous categorical variables and in addition exploits a specific structure of the age recorded in these datasets in order to create an additional blocking variable.  The ages recorded in these three datafiles present two gaps, this is, there are no homicides recorded in the 5--11 and 56--65 age intervals.  Thus, we create a new blocking variable that classifies ``kids'', ``young'', and ``elderly'' individuals.  We think it is safe to use this variable for blocking since no records with similar ages are assigned to different blocks.  Also, to help the EM algorithm to identify the appropriate clusters, we replaced $\hat P(S_p|\gamma^j)$ by 1 for those triplets with $\gamma^{j_f}_p=1$ for all the fields $f$ and for $p\in\{12/3, 13/2, 1/23, 123\}$.  This semi--supervised approach is a missing data problem under multinomial sampling \citep{DempsterLairdRubin77}.  We made the final assignments using nominal error levels $\mu_p=0.01$ for all $p$.

\begin{table}[hb]
 \vspace*{-6pt}
 \centering
\small{
  \begin{minipage}[b]{0.85\textwidth}
 \def\~{\hphantom{0}}
 \caption{Error rates of multiple record linkage assignments for Census (1) -- Forensics (2) -- Police (3) record triplets.  Three comparison data options for age of the victim and date of the homicide.  OME: Overall Misclassification Error, MWGE: Mean Within Group Error.}\label{t:agedateoptions}
  \begin{tabular*}{\columnwidth}{@{\extracolsep{\fill}}l@{\extracolsep{\fill}}l@{\extracolsep{\fill}}c@{\extracolsep{\fill}}c@{\extracolsep{\fill}}c@{\extracolsep{\fill}}c@{\extracolsep{\fill}}c@{\extracolsep{\fill}}c@{\extracolsep{\fill}}c}
\hline\hline\\[-8pt]
&  &  \multicolumn{5}{c}{Misclassification Error} \\
   \cline{3-7}\\ [-6pt]
& Age and Date Data & 1/2/3 & 12/3 & 13/2 & 1/23 & 123 & OME & MWGE \\
  \hline\\[-8pt]
 1. & Exact comparisons & 0.6203 & 0.2216 & 0.3915 & 0.0079 & 0.4444 & 0.5977 & 0.3371 \\
 2. & Three comparison  &  	&		&		&		&	&&	\\
    & categories & 0.0470 & 0.0109 & 0.0803 & 0.0510 & 0.0370 & 0.0471 &  0.0453 \\
 3. & Three comparison   &  	&		&		&		&	&&	\\
    & categories + blocking & 0.0365 & 0.0079 & 0.0598 & 0.0082 & 0.0370 & 0.0359 & 0.0299\\
    & Kid--Young--Elderly &    	&		&		&		&	&&	 \\\hline
\end{tabular*}\vskip5pt
\end{minipage}
}
\end{table}

In Table \ref{t:agedateoptions} we present different measures of the performance of the multiple record linkage decisions using the three different options for the inclusion of the information about age of the victim and date of the homicide.  These measures were obtained after comparing with the results of the hand matching procedure, which is thought to be more reliable.  Besides the usual misclassification errors, we present the mean within group error rate \citep{QiaoLiu09}, which controls the different sizes of the clusters $S_p$ by taking the average of the error rates for each $S_p$.  From the first age and date comparison data, we can see that the multiple record linkage procedure can produce catastrophic results if it is not used carefully.   For this scenario all the misclassification errors are very high, which indicates that the multiple record linkage process did not find the appropriate clusters.  For the first comparison data only exact comparisons were included, hence small differences in age and date were treated the same as large differences.  For the second age and date comparison data the results improved significantly.  The way these comparison variables were created is such that if there is exact agreement in age or in date, the three corresponding comparison variables agree.  If there is a difference of one unit, two of them agree, and if there is a difference of two units, only one of the variables agree.  This approach is more flexible to capture small measurement error in age and date.   The final approach additionally blocks three categories of age, which helps to reduce the number of misclassified triplets.  For this final approach all the measures of misclassification error are very close to zero, which indicates that multiple record linkage can provide good results if used properly.  Naturally, the good performance of the method depends on the specific datafiles to be linked and the models implemented.

We performed a bipartite record linkage for each of the three pairs of datafiles using the same blocking variables and the same comparison data as the third approach in Table \ref{t:agedateoptions}.  The assignments were also made using nominal error levels of $0.01$.  For the triplets on which a decision could be made, the overall misclassification error was 0.0435 and the mean within group error was 0.0311.  When trying to combine the decisions of the three independent procedures, however, we obtained a set of 43 record triplets on which we could not assign a decision.  Among this set of record triplets the multiple record linkage procedure coincided with the hand matching procedure in 32 cases (74\%).  Of course the performance of the method for those record triplets is not as good as the general performance, since these record triplets are usually the ones that are more difficult to classify.  However, multiple record linkage provides a decision along with a measure of uncertainty for that decision (namely, the matching probabilities), something that is not available from reconciling bipartite record linkages.

\section{SIMULATION STUDIES}\label{s:simulation}

In practice, the performance of our method will depend on several factors: (1) the amount of measurement error of the datafiles, (2) the number of common variables and their number of categories/variability, (3) the sizes of the datafiles and their overlaps, (4) the dependence structure among the recorded fields, (5) the existence of replicate records in the datafiles, etc.  Here we explore the performance of the proposed method under some simple scenarios, emphasizing how measurement error affects our results.  We used the R language to perform our simulations \citep{R10}.

\subsection{Generating Measurement Error}
\citet{TancrediLiseo11} use a simplified version of the \textit{hit-miss} model \citep{CopasHilton90} in order to generate measurement error.  This model for categorical information on records measured with error is given by
\begin{equation}\label{eq:hitmiss}
P(Y^{obs}_f=y_f^{c^o}|Y_f=y_f^{c})=(1-\beta_f) I(y_f^{c^o}=y_f^{c})+\beta_f/C_f
\end{equation}
where $Y_f^{obs}$ represents the observed field $f$ and $Y_f$ represents the true value of the field $f$.  Both $Y^{obs}_f$ and $Y_f$ have support $\{y_f^1,\dots,y_f^{c},\dots,y_f^{C_f}\}$, where $C_f$ represents the number of categories of the field $f$.  Equation (\ref{eq:hitmiss}) includes a measurement error parameter $\beta_f$ which represents the probability of measurement error for the field $f$. This model establishes that conditioning on the unobserved true values, we can model each single record field as a mixture of two components: the first component is concentrated on the true value while the second one is uniformly distributed over the support of the field \citep{TancrediLiseo11}.  In our simulation studies we do not generate error for the blocking variable.  For the numerical variables we generate error using the following model
\begin{equation}\label{eq:hitmiss1}
P(Y^{obs}_f=y_f^{c^o}|Y_f=y_f^{c})=(1-\beta_f) I(y_f^{c^o}=y_f^{c})+\beta_f\frac{2}{5}2^{-|y_f^{c^o}-y_f^{c}|}I(|y_f^{c^o}-y_f^{c}|<3),
\end{equation}
which allows measurement error around the true value.  For our simulation study we consider the same value of $\beta_f$ for all the fields subject to error (so we drop the subindex $f$).

\subsection{To Block or Not to Block?}

Blocking is usually an important component of record linkage since working with the complete cartesian product of the datafiles is computationally inefficient.  In this section we show that we need blocking to obtain good classification results.  Thus, we may want to block even in the presence of adequate computational power to handle the record linkage process on the complete cartesian product.

\begin{figure*}[th]
\centering
  \centerline{\includegraphics[width=0.85\linewidth]{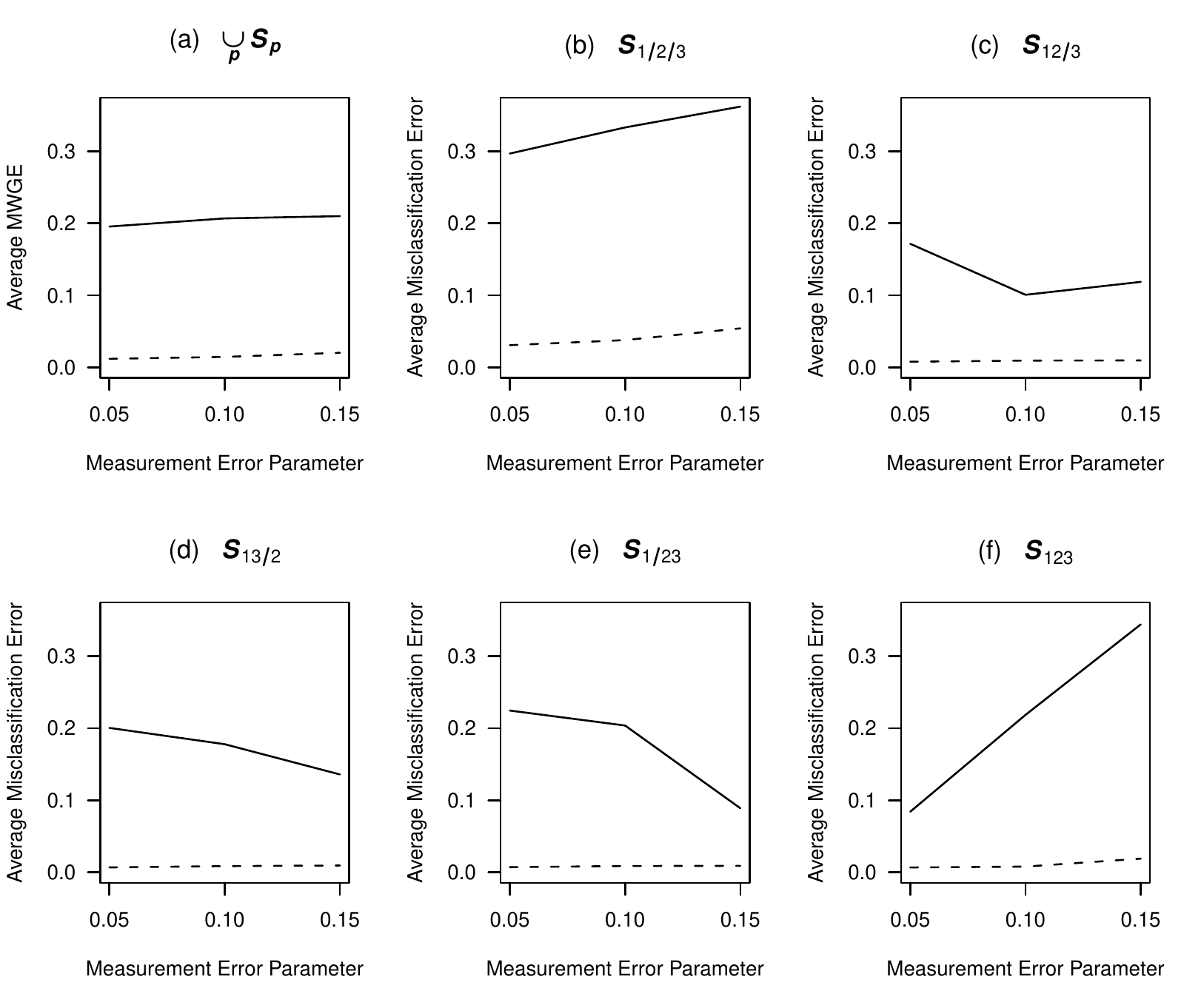}}
  \begin{minipage}[b]{0.85\textwidth}
  \caption{Measures of misclassification error for non--blocking (solid line) and blocking (dashed line) scenarios. }
\label{f:semisimresults}\end{minipage} 
\end{figure*}

We take the Census homicide data as the true population information and we generate three equal--size datafiles subject to measurement error.  We generate measurement error according to the model (\ref{eq:hitmiss1}) for date of the homicide and age of the victim.  We do not generate measurement error for sex of the victim and city of the homicide since we use these variables for blocking.  We simulate 100 triplets of datafiles and for each triplet we perform multiple record linkage using the second option of comparison data presented in Section \ref{s:application}.  In Figure \ref{f:semisimresults} we present the performance results for three values of the measurement error parameter: 0.05, 0.10, and 0.15.  We compare the results of our method without blocking (solid line) and after blocking by gender of the victim and city of the homicide (dashed line).  In panel (a) of Figure \ref{f:semisimresults} we average over all the simulations the mean within group error as a measure of the general performance of the method (or in other words, a measure of the performance of our method on $\bigcup_p S_p$).  In panels (b) to (f) we present the average misclassification error for each specific subset $S_p$.

We can see that, for this example, the effect of blocking is huge.  In general, the error rates are very large when we use no blocking, but they decay to values close to zero under blocking.  Note also that the larger the measurement error, the larger the error recovering the subsets $S_{123}$ and $S_{1/2/3}$, which indicates that measurement error causes true triple links to be missed and false links to be created.

\subsection{Number of Blocks and Low--Quality Fields}

In certain applications there are different blocking options and the possibility to include low--quality fields in the linkage process.  In this section we explore these scenarios.  We generate three databases containing five independent common fields across the different scenarios.  These first five fields contain 3, 5, 10, 10, and 15 categories, respectively, and each category is generated with equal probability.  We also use one additional independent blocking variable in order to check the performance of the method under blocking.  We consider three different blocking scenarios which correspond to 5, 10, and 15 categories of the blocking variable, where the categories are generated with equal probability.  For all the simulation scenarios, the sizes of the databases and their overlaps are the same as in the Colombian homicide data.

For one of the fields with 10 categories, we use $\beta=.7$ in order to simulate a scenario where a common variable is available, but it is known that its quality is low.  We keep $\beta=.7$ for the previous variable across three different measurement error scenarios for the remaining four fields.  These three scenarios correspond to three different values of $\beta$:  0.05, 0.10, and 0.15, and in each scenario the same $\beta$ is used to generate error for the remaining four fields.  Given the three true databases, we generate 100 triplets of observed databases using the hit-miss model (\ref{eq:hitmiss}).      For each triplet of databases we performed six implementations of the proposed methodology for multiple record linkage.  The six implementations correspond to the combination of including/excluding the low quality field and the three blocking options.  We made the final assignments using nominal error levels $\mu_p=.01$ for all $p$.

\begin{figure*}[th]
\centering
  \centerline{\includegraphics[width=0.85\linewidth]{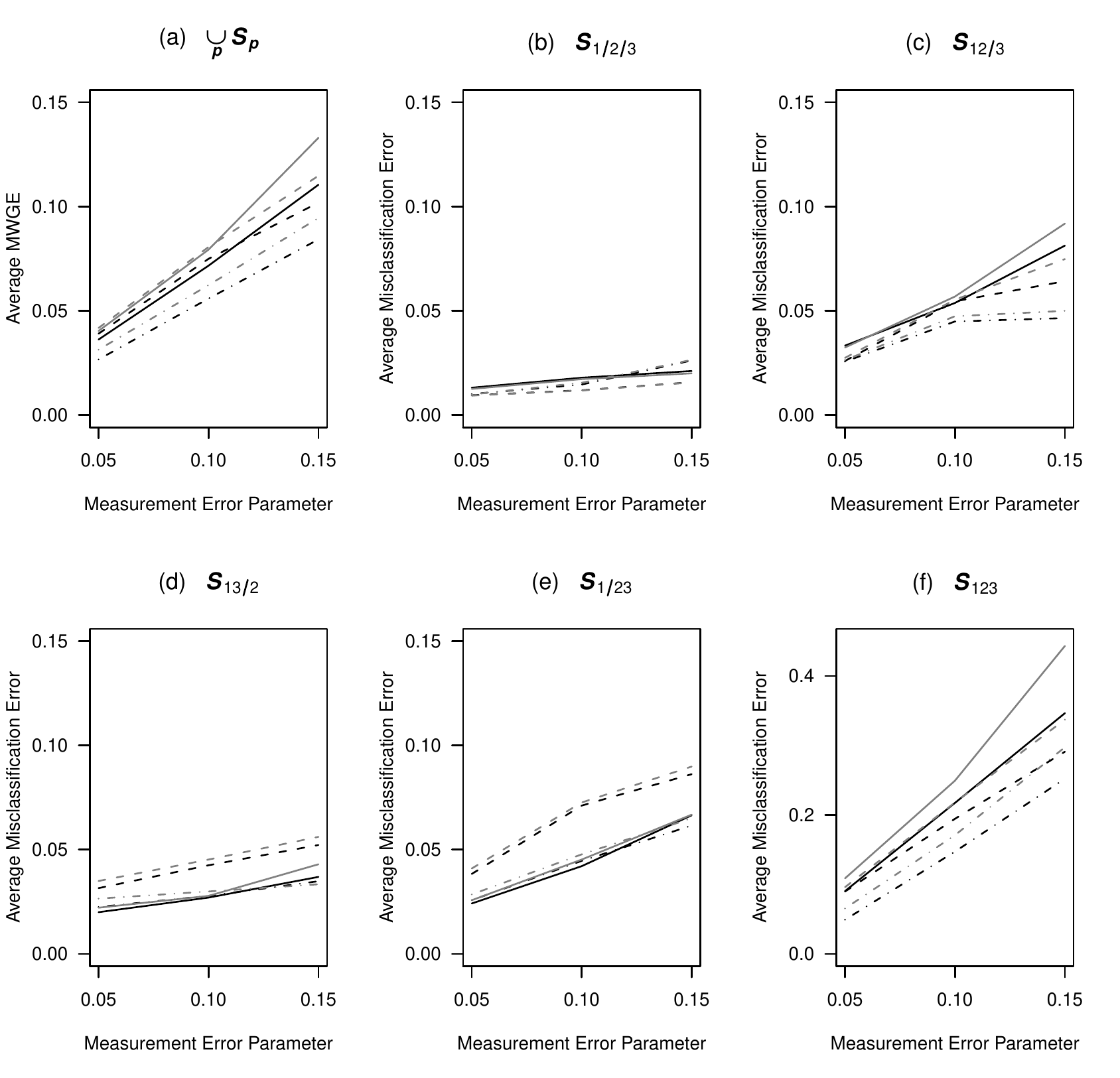}}
  \begin{minipage}[b]{0.85\textwidth}
  \caption{Measures of misclassification error for different number of blocks and inclusion/exclusion of low--quality fields.  The blocking scenarios are 5 blocks (solid line), 10 blocks (dashed line), and 15 blocks (dotdashed line).  The grey lines represent the performance of the method including the low--quality extra field. Note the different scale of panel (f).}
\label{f:simresults}\end{minipage} 
\end{figure*}

To evaluate the performance of the method in terms of recovering the classes $S_p$, we report the misclassification error rate for each class $S_p$ and the mean within group error rate \citep{QiaoLiu09} for the triplets that were assigned to a certain group.  The mean within group error rate is more meaningful than the overall misclassification error for record linkage since the groups $S_p$ are extremely unbalanced, e.g., the subset $S_{1/2/3}$ is massive whereas the subset $S_{123}$ is extremely small.  We present the results in Figure \ref{f:simresults}, where panel (a) shows the average over all the simulations of the mean within group error (MWGE) and panels (b) to (f) show the average misclassification error for each class $S_p$.  All the panels show the performance measures as a function of the measurement error parameter.  The solid, dashed, and dotdashed lines represent the error values for the method with 5, 10, and 15 blocks, respectively.  The grey lines represent the method including the low--quality extra field.  Note that the scale of the vertical axes is the same for panels (a) to (e), but we present panel (f) with a different scale since the errors for the subset $S_{123}$ are significantly larger compared to the other subsets.

We can see that, in general, the larger the measurement error, the larger the error rates, which is something that one would expect.  We can also see that under all the scenarios, increasing the amount of blocking decreases the error rates.  In particular, note in panel (f) that blocking has a huge impact on the reduction of the misclassification for the class $S_{123}$.  Finally, we note that for each blocking scenario, the inclusion of the low--quality extra field increases the error rates.

\section{CONCLUSIONS AND FUTURE WORK}

Our method provides a framework for the integration of more than two datafiles without common identifiers.  The ideas are an extension of the theory proposed by \citet{FellegiSunter69} and its more modern implementations, as in \citet{Winkler88} and \citet{Jaro89}.  The method solves the problem of obtaining non--transitive decisions, as it is common when reconciling bipartite record linkages.  Our method also provides matching probabilities for the record $K$-tuples, something that is not available from reconciling bipartite record linkages, but that is necessary in order to incorporate the uncertainty of the linkage procedure in posterior analysis such as regression \citep{LahiriLarsen05}.  We proposed a decision rule which is optimal under the availability of the true matching probabilities.  In practice, however, the optimality of the decision rule hinges on the availability of well-calibrated probability models, i.e.,  good estimates of the probability of a particular $K$-tuple belonging to the subsets $S_p$.  Thus, we need to consider models that go beyond the present one and that capture dependencies between fields \citep[e.g., see][]{LarsenRubin01}.  Nevertheless, even using a naive model, our method performed well both in the integration of the Colombian homicide datafiles and in our simulations.

We believe our method holds promise in the context of record linkage for
    census coverage measurement evaluation.  For example, the U.S.  Census
Bureau has for several decades done a two--sample linkage between the
actual enumeration and data from a post-enumeration survey based on data
from a nationwide sample of census blocks \citep{Hogan92,Hogan93}.
Additional sources of data that could be used to improve coverage estimation include the American Community Survey and various
administrative record files.  Incorporation of them  would require linkage of
$K \ge 3$ datafiles, using methods that could build upon the work
described here that would take into account multiple sampling designs
and census adjustments such as imputations and erroneous enumerations.

\appendix
\numberwithin{equation}{section}
\section{APPENDIX: PROOFS}\label{ap:optimrule}

In the proofs presented below we use the notation introduced in Section \ref{s:matchprobs}, where for instance, $P(S^c_{p})$ means  $P(r_j\in S^c_{p})$, and so on.

\textit{Proof of Proposition 1.}  The ordering of $w_{p}^j$ is the same as the ordering of $\text{logit}\left[ P(S_{p}|\gamma^j)\right]$ since
\begin{eqnarray*}
w_{p}^j &=& \log\frac{ P(S_{p}|\gamma^j)/P(S_p)}{ P(S^c_{p}|\gamma^j)/P(S^c_{p})}\\
&\propto& \log\frac{ P(S_{p}|\gamma^j)}{ P(S^c_{p}|\gamma^j)}\\
&=& \text{logit}\left[ P(S_{p}|\gamma^j)\right].
\end{eqnarray*}
Finally, the logit function is a monotonic increasing function of its argument, thus the ordering of $\text{logit}\left[ P(S_{p}|\gamma^j)\right]$ is the same as the ordering of $ P(S_{p}|\gamma^j)$.
\\

\textit{Proof of Theorem 1. Optimality of the Generalized Fellegi--Sunter Linkage Rule.}

Let us define the set of possible decisions for a record $K$-tuple.  Let us call $D_p$ the decision of assigning a record $K$-tuple to the subset $S_p$ and $D_u$ the decision to keep the record $K$-tuple undeclared.  Thus, a decision function $d$ is a $(B_K+1)$-tuple given by
\begin{equation*}
d(\gamma^j)=\left(P(D_{1/2/\dots/K}|\gamma^j),\dots,P(D_{p}|\gamma^j),\dots,P(D_{12\dots K}|\gamma^j),P(D_u|\gamma^j)\right)
\end{equation*}
where
\begin{equation*}
P(D_u|\gamma^j)+\sum_{p\in\mathbb{P}_K}P(D_{p}|\gamma^j)=1 .
\end{equation*}
The proposed decision rule $L_0$ is such that
\begin{equation*}
              \begin{array}{ll}
                P_0(D_{p}|\gamma^j)=1, & \text{ if \ }  (j)_p\leq(j')_p-1; \\
                P_0(D_{u}|\gamma^j)=1, & \text{ if \ }  (j)_p\geq(j')_p; \\
              \end{array}
\end{equation*}
for $(j)_p$ in the subset of record $K$-tuples for which $P(S_{p}|\gamma^j)$ is maximum and  $(j')_p$ is obtained as in the statement of Theorem 1.  This decision rule minimizes the probability of assigning each record $K$-tuple to the wrong subset $S_p$ or keeping it undeclared, subject to a set of admissible error levels $\mu_p=P(D_p|S_p^c)$, $p\in\mathbb{P}_K$.  For decision rules $L_0$ and $L_1$
$$
\mu_p=P(D_p|S_p^c)=\sum_{(j)_p} P_0(D_{p}|\gamma^{(j)_p}) P(\gamma^{(j)_p}| S^c_{p})=\sum_{(j)_p} P_1(D_{p}|\gamma^{(j)_p}) P(\gamma^{(j)_p}| S^c_{p}).
$$
From the construction of $L_0$ we obtain
$$
\sum_{(j)_p\leq(j')_p-1} P(\gamma^{(j)_p}| S^c_{p})=\sum_{(j)_p} P_1(D_{p}|\gamma^{(j)_p}) P(\gamma^{(j)_p}| S^c_{p})
$$
or
\begin{equation}\label{eq:proof1a}
\sum_{(j)_p\leq(j')_p-1} P(\gamma^{(j)_p}| S^c_{p})\left[1-P_1(D_{p}|\gamma^{(j)_p})\right]=\sum_{(j)_p\geq(j')_p} P_1(D_{p}|\gamma^{(j)_p}) P(\gamma^{(j)_p}| S^c_{p}).
\end{equation}
Since
$$
P(\gamma^{(i)_p}| S_{p})P(\gamma^{(j)_p}| S^c_{p})\leq P(\gamma^{(j)_p}| S_{p})P(\gamma^{(i)_p}| S^c_{p})
$$
whenever $(j)_p<(i)_p$ we have
\begin{eqnarray}
&&\left[\sum_{(j)_p\geq(j')_p} P_1(D_{p}|\gamma^{(j)_p}) P(\gamma^{(j)_p}| S_{p})\right]\left[\sum_{(j)_p\leq(j')_p-1} P(\gamma^{(j)_p}| S^c_{p})\left[1-P_1(D_{p}|\gamma^{(j)_p})\right]\right]\nonumber \\
&& \leq  \left[\sum_{(j)_p\geq(j')_p} P_1(D_{p}|\gamma^{(j)_p}) P(\gamma^{(j)_p}| S^c_{p})\right]\left[\sum_{(j)_p\leq(j')_p-1} P(\gamma^{(j)_p}| S_{p})\left[1-P_1(D_{p}|\gamma^{(j)_p})\right]\right];\label{eq:proof1b}
\end{eqnarray}
dividing (\ref{eq:proof1b}) by (\ref{eq:proof1a}) we obtain
\begin{equation*}
\left[\sum_{(j)_p\geq(j')_p} P(\gamma^{(j)_p}| S_{p})P_1(D_{p}|\gamma^{(j)_p}) \right] \leq  \left[\sum_{(j)_p\leq(j')_p-1} P(\gamma^{(j)_p}| S_{p})\left[1-P_1(D_{p}|\gamma^{(j)_p})\right]\right]
\end{equation*}
from which
\begin{equation*}
\left[\sum_{(j)_p} P(\gamma^{(j)_p}| S_{p})P_1(D_{p}|\gamma^{(j)_p}) \right] \leq  \left[\sum_{(j)_p} P(\gamma^{(j)_p}| S_{p})P_0(D_{p}|\gamma^{(j)_p})\right],
\end{equation*}
which is the same as
$$
P_1(D_p|S_p)\leq P_0(D_p|S_p),
$$
which implies
\begin{equation}\label{eq:optimalgivenSp}
P_1(D^c_p|S_p)\geq P_0(D^c_p|S_p)
\end{equation}
for all $p\in\mathbb{P}_K$.  Note that the probability of taking a wrong decision or not deciding can be written as
$$
\sum_{p\in\mathbb{P}_K} P(D_p^c\cap S_p)=\sum_{p\in\mathbb{P}_K} P(D_p^c|S_p)P(S_p),
$$
which is minimized by the generalized Fellegi--Sunter linkage rule $L_0$, as we can see using (\ref{eq:optimalgivenSp}).

\bibliographystyle{apalike}
\bibliography{biblio}

\end{document}